\documentstyle[preprint,aps]{revtex}
\begin{document}
\title{The Bose--Hubbard Model and Superfluid Staircases in $^{4}$He Films}
\author{G.T. Zimanyi$^{1}$, P.A. Crowell$^{2}$, R.T. Scalettar$^{1}$,
and G.G. Batrouni$^{3}$}
\address{$^{1}$Department of Physics, University of California, Davis, CA
95616}
\address{$^{2}$Laboratory of Atomic and Solid State Physics and the Materials
Science Center, Cornell University, Ithaca, NY 14853-2501}
\address{$^{3}$Thinking Machines Corporation, 245 First Street, Cambridge,
MA 02142}
\date{\today}
\maketitle
\begin{abstract}
The recent experimental observation of step-like structure
in the superfluid density of $^{4}$He films on graphite
is interpreted in terms of passage close to Mott insulating
lobes in the Bose--Hubbard model. Plateaus develop in the
superfluid density
near the completion of each layer because of the He-He
repulsion.  We modify the Bose--Hubbard Hamiltonian,
using a density-dependent He-He interaction term
to model the He-substrate potential.
Quantum Monte Carlo simulations yield
results which agree well with the experimental data.
\end{abstract}
\newpage
%\begin{narrowtext}
$^4$He films have been valuable in studying
quantum-mechanical effects in systems of reduced dimensionality\cite{somebook}.
In particular, there has been a large body of work on superfluidity
in thin films\cite {Reppy}. Most of these experiments have
been conducted on disordered substrates.
%Surprisingly, however, superfluidity
%in $^4$He films adsorbed on {\it ordered} substrates
%has received much less experimental attention,
A recent Letter\cite {Crowell1}, however, reported on the measurements of the
superfluid density in $^4$He films
adsorbed on the basal plane of graphite, which is ordered on
atomic length scales.
A number of novel features were observed, including the appearance of
superfluidity in a narrow range of coverage in the second layer,
and a subsequent staircase structure in the superfluid signal as a function
of coverage. The plateaus in this staircase appear
near completion of the third to sixth layers.  This discovery
complemented the vapor pressure and third sound measurements
of Zimmerli {\it et al.}\cite {Zimmerli},
 which showed that the chemical potential $\mu$ also grows
in steps out to the completion of the seventh layer and that the third
sound velocity has structure that is periodic in coverage.
The steps in $\mu$ occur at layer completion, where the
superfluid signal as measured in the torsional oscillator experiment\cite
{Crowell1} is nearly constant.

Meanwhile, a body of analytic and
numerical work
\cite{Fisher,Batrouni1,Scalettar1,Trivedi1,Trivedi2,Rokhsar,Runge,Kampf,Freericks}
on models of interacting bosons, with and without disorder, has established
the corresponding phase diagrams and the critical properties of these systems.
The ``Bose--Hubbard'' (BH) Hamiltonian
\begin{equation}
H=-t\sum_{<ij>} (a_{i}^{\dagger}a_{j} + a_{j}^{  - \mu \sum_{i} n_{i} + V
\sum_{i} n_{i}^{2}
\label {eq:eq1}
\end{equation}
is believed to represent much of the qualitative physics
of the helium system, and upon incorporating disorder,
for the related problem
of granular superconductors\cite{Fisher}.
Here $a_{i}$ is a boson destruction operator at site $i$, and
$n_{i}=a_{i}^{\dagger}a_{i}$.  The transfer integral $t$ sets the scale
of the kinetic energy, while $\mu$ and $V$ are
the chemical potential and the boson--boson repulsion.

In this paper we suggest analogies
between the dependence of $\mu(n)$ and $\rho_s(n)$ on the density $n$
in $^{4}$He films adsorbed on graphite and in the BH model.
It already has been established
that the BH model exhibits a step--like structure
in the chemical potential as a function of $n$.
To understand this,
consider the empty lattice for the case $t=0$.
Particles can be added to the system at no cost, i.e. $\mu=0$, until
each site is occupied by a single atom. Once each site is occupied,
however, additional atoms must be given an energy $V$ to overcome
the repulsion of the atom which is already present,
Continuing this argument, we see that the chemical potential develops a
jump, which is proportional to
$V$ at the commensurate densities $n=1,2,3, ...$ atoms/site
and is constant for all intermediate densities.
These jumps in $\mu$ also imply the appearance of a gap $\Delta$ in
the spectrum.
For the case $t>0$, the steps decrease in size, and the chemical
potential is no longer constant at incommensurate fillings, but
the step--structure disappears only gradually with increasing~$t$.

The presence of the gap for sufficiently small
$t/V$, i.e. large interaction strengths,
means that the interparticle repulsion $V$
{\it localizes} the bosons into Mott-type insulating phases at
commensurate fillings. Inside these phases the superfluid de$\rho_{s}$
vanishes.  A critical hopping strength $t_c~\sim~V$ is
required for delocalization, giving rise to superfluidity.
Away from commensuration the bosons are delocalized at all hopping strength,
and of course at low temperatures they form a superfluid.
The qualitative phase diagram for clean systems
is shown in Fig.~1.\cite{Fisher}
This phase diagram has recently been confirmed quantitatively by
Monte Carlo simulations,\cite{Scalettar1,Trivedi1}
by coarse graining methods,\cite{Kampf} and by low order
perturbation expansions.\cite{Freericks}
When disorder is incorporated in the model,
various glassy phases have been proposed
which are characterized by a vanishing $\rho_{s}$
and $\Delta$,\cite{Fisher,Batrouni1,Trivedi2,Rokhsar,Ma}
but their exact location on the phase diagram is not yet established.

The staircase structure in the superfluid density $\rho_s$ can also
be understood in the context of the BH model. As discussed above,
the superfluid density vanishes inside and at the boundaries of the Mott
insulating lobes in Fig.~1.
The superfluid density is expected to grow smoothly with
$\mu$ when the system is deep in the superfluid part of the phase diagram.
In Fig.~1 we show contours of constant $\rho_s$ using dotted curves.
%The system follows some path in the parameter space as the density
%and $\mu$ are increased.
As the density is increased, the system follows a nearly vertical
path in parameter space
similar to that shown by the dashed curve in Fig.~1. The choice of this
path will be discussed below.
If this trajectory passes in the vicinity of the Mott lobes, it will be
nearly tangential to the contours, resulting in plateaus in $\rho_s$ vs. $n$.

The BH model is only an approximation of a real $^4$He film.  First,
it does not incorporate the long-range attractive tail
of the He-He potential.  As a consequence, notransitions\cite {Bretz,Greywall}
that occur in the first two layers of adsorbed $^4$He appear in the BH phase
diagram.
As we are interested in superfluid properties, and the first two layers
are known to solidify\cite{Greywall,Lauter}, we will only study the system at
coverages above two layers.
We consider the first two layers to form a pseudo-substrate for
the overlying superfluid film, and we use the
BH Hamiltonian only to model layers added above the
two solidified layers.
Therefore the first layer in our calculations
will correspond to the third experimental layer.
A second shortcoming of the BH model, which we address here,
is the exclusion of the He-substrate potential.
The two leading approaches to the He-substrate interaction
are the Dzyaloshinskii-Lifshitz-Pitaevskii (DLP) \cite{dlp} and
Frenkel-Halsey-Hill (FHH) \cite{fhh} theories. Both
model the fluid and the substrate as structureless continua,
disregarding the effects of layering as well as structural phase
transitions in the film.
They consider the shift in the chemical potential, $\Delta\mu $,
caused by the presence of the smooth surface.
In both cases the result can be cast in the form:
\begin{equation}
\Delta \mu(d) = - \gamma(d)/ d^3,
\label{eq:eq2}
\end{equation}
where the coefficient $\gamma(d)$ is a function of the layer thickness $d$.
This form can be derived by assuming a van-der-Waals interaction between
the helium and the surface. The FHH theory adopts the simplest assumption:
$\gamma(d)= \gamma(0) $, a constant.

DLP provide a generally formulated many-body approach.
As shown by Cheng and Cole \cite{Cole},
the FHH formula is the first term upon expanding the DLP theory in the limit
of small densities and polarizability, neglecting retardation.
The higher order terms of the DLP expansion can be neglected at the
thicknesses under considerOn the other hand, Zimmerli {\it et al.} have shown
that the FHH formula overestimates the layer thickness by nearly a full
layer\cite{Zimmerli}.
A reasonable modification has been proposed by Cheng and
Cole\cite{Cole2}, who treat the solid layers as a pseudosubstrate,
leading to a modified $\gamma(d)$:
\begin{equation}
\gamma (d') = 1772 K\AA ^3 ~ [0.07 + (1+D/(d'))^{-3}] ~ ,
\label{eq:eq3}
\end{equation}
where $D=4.42$ \AA\
is the thickness of the first two solid layers and $d'=d-D$ is the thickness
of the fluid film.
The parameters used in Eq.~3 were extracted by Cheng and Cole
from experiments\cite{Cole2}.
Measurements of the third sound velocity by Zimmerli
{\it et al.} \cite{Zimmerli} provided quantitative support for this picture.
We will use Eqs.~\ref{eq:eq2} and~\ref{eq:eq3} to represent
the effects of the substrate and the two solid layers on the chemical
potential of films thicker than two atomic layers.

We solved for the equilibrium thermodynamic properties of the
BH Hamiltonian using the Quantum Monte Carlo (QMC)\cite{Batrouni2} technique.
This approach evaluates operator expectation values, treating the
many body correlations in an exact manner by computing the
imaginary time evolution operator $e^{-\beta H}$ using
stochastic techniques.  The results presented here are for
$8 \times 8$ spatial lattices with a discretization
of imaginary time equal to $1/16$ of the bandwidth.
Convergence of the results and finite size effects have
been carefully checked.

Our Quantum Monte Carlo calculation is conducted in the canonical
ensemble, so that $\mu$ is not an independent variable.
We therefore model the density dependence of the chemical potential
by a density-dependent interaction term $V(n)$,
where $n=<n_{i}>$ is the density in $^4$He atoms/site. The underlying idea is
that the interaction term in the Ha$H_{int} ~ = ~ \sum _{i} V(n) n_{i}^2 ~
\approx ~ \sum _{i} V(n) n ~ n_{i} ~ ,$
can be thought of as renormalizing $\mu$. We use the relation
\begin{equation}
V(n) = V + \Delta \mu (d(n))/n =
V - C~[{1\over n^4} + {0.07\over n (n-2)^3}]~ ,
\end{equation}
valid for densities greater than 2 atoms/site.
We regard both $V$ and $C$ as phenomenological
parameters.  Their values are determined
by requiring that the path followed in the $\mu/V-t/V$ plane produce
the observed plateau structure in the superfluid density.
We emphasize that when the full chemical potential is extracted
from the ground state energy $E_G$, via $\mu = \partial E_G/\partial n$,
this smooth shift will be further renormalized by the effects of
the He--He interaction, leading to the staircase structure.

We compare the results of our calculation
with the torsional oscillator
measurements of Crowell and Reppy\cite{Crowell1,Crowell2}.
Their experimental apparatus comprises a cell containing exfoliated graphite
mounted on a torsion rod.  The measurements discussed here
were made while adding $^4$He to the cell at constant temperature.
For each coverage, the superfluid
period shift $\Delta P$ was determined by taking
the absolute value of the difference between the resonant period and
the period that would have been measured if the film were non-superfluid
and hence locked to the substrate. The period shift $\Delta P$ is
proportional to the superfluid density $\rho_s$ if the film is a uniform
fluid.  The experimental technique is discussed in greater
detail elsewhere\cite {Crowell3}.

In Figure~2, we show the results of a sweep on a path parametrized
by $V/t=8.5$ and $C/t=88$.
An estimate of the energy scale, $t\sim 1.5$K, was obtained by
Fisher and Liu\cite{Fisherliu} by fitting the phase diagram
of a lattice model of bulk helium to the experimental In our calculation $\beta
t=4$,
implying a temperature $T\sim 400$ mK.
The experimental data at 500~mK, scaled by a constant, are also shown in
Fig.~2, which displays the central result of this paper, namely
the close reproduction of the plateau structure of $\rho_{s}$ in the
experiments
by quantum simulations of the Bose--Hubbard model.
The agreement of the calculation with the experimental data
suggests that the origin of the plateaus in $\rho_s$ is the localizing
effect of the interparticle repulsion.
The plateaus in $\rho_s$ are a robust feature of the model, appearing over
a wide range of possible paths, including $V(n)=V$, a constant.
The quantitative agreement is best, however, for the trajectory that
follows from Eq.~4. The dip in the numerical calculation near n=3 is due the
close proximity of the trajectory to the Mott lobe, but this feature
is sensitive to the choice of parameters.

Additional evidence supporting our conclusions can be found
in the experimental data at 20~mK, which are shown along with the 500~mK
data in Figure~3.  The plateau at $n=4$ is even more pronounced
with decreasing T. This observation is consistent with the fact that
the localizing effects of the repulsion are enhanced at
lower temperatures\cite{Batrouni1,Kampf}. We did not carry out numerical
simulations at 20 ~mK because of the prohibitively long computation time.

A plateau at the third layer may follow, however, from
structural phase transitions in the film.
Lauter {\it et al.} \cite{Lauter}, using neutron scattering,
have observed a sudden increase in the density of
the second layer incommensurate solid in the vicinity
of third layer completion.
The atoms that go into the second-layer solid
in this case come at the expense of the fourth layer, which does not start
to fill until the reconstruction of the second layer
is complete.  The expeare consistent with this hypothesis, since there is no
indication that
the atoms added to the film between 28 and 30 atoms/nm$^2$ make
a contribution to the superfluid signal.  The plateau near third-layer
completion is also
much narrower than those found at higher coverages and its onset occurs
closer to layer completion, suggesting that the underlying mechanism
may be different from
that responsible for the behavior observed at higher coverages.
If the reconstruction occurs, the coverage at third layer completion should
be shifted upward from the conventional value of 28.0 atoms/nm$^2$.
We have done so in Fig.~2, finding the best agreement between theory
and experiment when the fourth layer starts filling at 29.3 atoms/nm$^2$.
This shift is somewhat greater than the lower bound of 0.8 atoms/nm$^2$
inferred
from the neutron scattering data of Lauter {\it et al.}\cite {Lauter}, but
it is less than $\sim 30\%$ of the width of the plateau at fourth layer
completion.

An alternative scenario for the plateaus has been suggested
by Clements {\it et al.}\cite{Clements1}.
They find that liquid clusters form at low densities in the third, fourth
and fifth layers.  These do not contribute
to the superfluid signal until they percolate across a macroscopic region of
the substrate.  As a result there is a plateau in $\rho_{s}$ at the beginning
of each layer, followed
by a rapid increase once percolation occurs.
The clustering theory predicts the appearance of plateaus
{\it after} layer completion. However, the plateaus in the experimental
data corresponding to the
fourth, fifth and sixth layers start several atoms/nm$^2$ {\it before}
layer completion.
We emphasize that even without any reconstruction shift, our
numerical simulations
yield plateaus which begin substantially before layer completion, in
agreement with the experiment.

In summarof plateaus in the superfluid density of thin films of $^4$He on
graphite.
Although
other mechanisms may be important at lower coverages, we find that
the formation of plateaus near the completion of the fourth and higher
layers is driven primarily by the strong He-He
repulsion.
Our simulations, which incorporate the He-substrate potential in an
approximate way, agree quantitatively with the experimental data.
Improvements on the model, such as accounting
for the long-range attractive part of the He-He interaction, may
enhance the agreement with the data at lower densities.

We want to thank H.~Godfrin, J.D.~Reppy, and  M.~Cole for useful discussions,
and B.E.~Clements and coworkers for providing us with a preprint.
P.A.C is grateful to AT\&T Bell Laboratories for fellowship support.
This work has been supported by NSF grants DMR 92-06023 (Davis)
and  DMR 89-21733 (Cornell).  The work at Cornell was also supported
by the MRL program of the NSF through Award No. DMR-9121654; MSC Report
No.~7655.

\newpage

{\bf Figure Captions}

Fig.~1: Qualitative phase diagram of the Bose-Hubbard model.
The solid lines are the boundaries of the localized Mott lobes.
The dotted curves are
contours of constant superfluid density, and the trajectory followed by
the {\it modified} system upon increasing the density is shown as a
dashed curve.

Fig.~2.  The superfluid density $\rho_s$ determined from torsional
oscillator measurements (circles) and the quantum Monte Carlo
simulation (triangles).  The experimental data has been scaled
by a constant and its coverage scale has been shifted as described in the text.

Fig.~3.  The period shift data at 20~mK (closed circles) and 500~mK (open
circles) are shown as a function of the $^4$He coverage.  The dashed lines
indicate layer completion according to the coverage scale used in
Fig.~2.

%\end{narrowtext}
\end{document}